\documentclass[12pt,letterpaper,final]{article}
\usepackage[latin1]{inputenc}
\usepackage{amsmath}
\usepackage{amsfonts}
\usepackage{amssymb}
\usepackage{graphicx}
\usepackage{nicefrac}
\usepackage{authblk}

\newcommand{\kket}[1]{|{#1}\rangle}

\newcommand{\ql}{\textquotedblleft}

 \newcommand{\summation}[2]{\displaystyle\sum\limits_{{#1}}^{{#2}}}
\input{Qcircuit}

\author[1,2]{Ben Criger}
\author[1,2]{Daniel Park}
\author[1,2,3]{Jonathan Baugh}

\affil[1]{Institute for Quantum Computing, University of Waterloo, Waterloo, Ontario, N2L 3G1, Canada}
\affil[2]{Department of Physics and Astronomy, University of Waterloo, Waterloo, Ontario, N2L 3G1, Canada}
\affil[3]{Department of Chemistry, University of Waterloo, Waterloo, Ontario, N2L 3G1, Canada}

\title{Few-Qubit Magnetic Resonance Quantum Information Processors: Simulating Chemistry and Physics}
\date{\today}
\begin{document}
\maketitle
\tableofcontents
\newpage
\section{Nuclear Magnetic Resonance QIP}
\subsection{Introduction}

Quantum information science is an interdisciplinary area of study combining computer science, physics, mathematics and engineering, the main aim of which is to perform exponentially faster computation using systems which are governed by the laws of quantum mechanics. The task of performing quantum algorithms or simulations is referred to as a quantum information processing (QIP). More fundamentally, quantum information science provides a new lens through which to view quantum physics. An interesting consequence of this new perspective is that any system which possesses an accurate description in terms of quantum mechanics may be simulated by any other quantum system of similar size~\cite{Feynman}. Such simulations provide a link between chemistry and quantum information science, through the discipline of quantum chemistry. 

In order to accurately predict the results of an atomic or molecular interaction, it is sometimes necessary to formulate a Hamiltonian model for the molecules in question. The Hamiltonian assigns energy values to certain states of the physical system, called basis states. This model is then incorporated into an appropriate equation of motion; either the Schr\"odinger equation (for the non-relativistic case) or the Dirac equation (for the relativistic case). The number of basis states grows exponentially with the physical size of the system, making the solution of the equation of motion a difficult task for a classical computer. 

These interactions can be efficiently simulated on a quantum computer, however, since a quantum operation is effectively performed in parallel across all basis states taken into a superposition.  Quantum simulations of the type described in this chapter are thought to comprise a set of attainable milestones for quantum information processing, given that there exist relatively simple quantum systems whose dynamics are not easily simulated using classical computers (for example, the 10-body Schr\"odinger equation for electrons in a water molecule ~\cite{ChanHeadGordon}. 

The remainder of this chapter is organized as follows: We first detail a series of quantum algorithms which are useful in simulation of chemical phenomena, then describe a class of QIP implementations using nuclear magnetic resonance (NMR) and electron spin resonance (ESR). In conclusion, we discuss some recent experiments and progress toward scalable spin-based implementations. 

\subsection{Quantum Algorithms for Chemistry}
The ostensible goal of a chemical simulation is to extract a small amount of data about a given process; a rate constant, ground state energy, or other quantity of interest. However, it is often necessary to manipulate a large data structure, such as a full molecular electronic configuration, in order to obtain the desired output. This is the exact class of mathematical problem for which a quantum computer is thought to be superior ~\cite{Feynman,Lloyd}, often qualified with the notion that using a quantum system to simulate another quantum system provides a physically elegant intuition. In this section, we describe how quantum information processing can provide the desired data, without an exponential growth in the required computational resources. The algorithms used include straightforward digital simulation of Hamiltonian mechanics using the Trotter expansion, as well as implementations of the adiabatic algorithm to simulate ground state properties of a large class of physical Hamiltonians. These comprise the vast majority of proposed simulation techniques; though they are far from a complete list of QIP architectures. 
\subsubsection{Digital Quantum Simulation}
In order to simulate continuous degrees of freedom, such as position and momentum, it is often necessary to \textit{discretize} these degrees of freedom onto a finite space, to ensure that the amount of memory required to store their values is bounded, while also ensuring that the resolution remains high enough to mimic the dynamics of the continuous-variable system under consideration. This method was pioneered by Zalka~\cite{ZalkaPRSL} and Wiesner~\cite{Wiesner}, to simulate Hamiltonians of the form
\begin{equation}
\hat H = \dfrac{\hat p^2}{2m} + V(\hat x).
\end{equation}
To accomplish this, position is discretized as detailed above, and encoded into a quantum register. In this way, an arbitrary superposition of $2^m \,\, \hat x$-eigenstates can be stored in $m$ qubits.

To simulate the evolution under this Hamiltonian, it is useful to perform a Trotter decomposition to first order~\cite{Trotter1959}:
\begin{equation}
\hat U_{\mathrm{evol}}=\exp{\left[-i\left(\dfrac{\hat p^2}{2m} + V(\hat x)\right)t\right]} \approx \exp{\left[-i\dfrac{\hat p^2}{2m}t\right]}\exp{\left[-i V(\hat x)t\right]}.
\end{equation}
This approximation, valid for small $t$, expresses the evolution under the Hamiltonian to be simulated in terms of operators which are diagonalized in known bases (momentum and position, respectively). Since $\hat V$ depends only on $x$, it is diagonal in the basis selected above. Assuming that diagonal operators can be implemented quickly on a quantum computer, all that remains is to find an efficient quantum circuit that will transform operators which are diagonal in the $x$-basis into operators which are diagonal in the $p$-basis. The quantum Fourier transform fulfils this requirement, and has been widely studied~\cite{WeinsteinRef15,WeinsteinRef16} and implemented~\cite{Weinstein2001,ChiaveriniBritton,ScullyPRA02}. The effective evolution operator, for a single particle in one dimension, is then:
\begin{equation}
QFT^{\dagger}\exp{\left[-i\dfrac{\hat D_{p^2}}{2m}t\right]}QFT\exp{\left[-i V(\hat x)t\right]},
\end{equation} 
where $\hat D_{p^2}$ is the diagonal operator whose entries are the eigenvalues of $\hat p ^2$. 

Using $n$ $m$-qubit registers, one can simulate a system of $n$ interacting degrees of freedom. This simulation involves the class of Hamiltonians 
\begin{equation}
\hat H = \sum_j \dfrac{\hat p_j^2}{2m_j} + V(\hat x_1,\hat x_2,\ldots \hat x_j).
\end{equation}
Here, the index $j$ lists the degrees of freedom in the Hamiltonian, which may correspond to distinct particles or dimensions. The Trotter expansion of the evolution under this Hamiltonian requires one quantum Fourier transform per degree of freedom. 

In order to benefit from the efficiency offered by quantum simulation, it is also necessary to ensure that measurement can be achieved in polynomial time. Kassal et al.~\cite{KassalPNAS} demonstrated that reaction probabilities, state-to-state transition probabilities and rate constants can be extracted from a quantum simulation of a chemical process in polynomial time. To accomplish these measurements, the classical algorithm for generating a transition state dividing surface is invoked, subdividing the simulation space into regions corresponding to products and reactants. Measurement of the single bit corresponding to the presence of the wavefunction in the reactant or product regions can be used in conjunction with phase estimation~\cite{KnillOrtizSomma}, so that the precision of the transition probability scales as $N^{-1}$, where $N$ is the number of single-bit measurements.

Using a limited capacity for quantum control, it is possible to simulate quantum dynamics, with immediate applications to small numbers of reacting atoms. Note that no Born-Oppenheimer approximation has been used. Indeed, the quantum algorithm is more efficient without this approximation, because the Born-Oppenheimer approximation requires the calculation of potential energy surfaces at many points throughout a given simulation. The realization of chemical simulations by the algorithm above, while scalable, requires hundreds of qubits in order to simulate a few particles with sufficient spatial precision, and $\sim 10^{12}$ elementary quantum gates to obtain sufficient precision in time.

\subsubsection{Adiabatic Quantum Simulation}
Given that the number of quantum gates used in the simulation detailed above is prohibitive for implementation in the near future, it is preferable to search for alternatives to gate-based quantum computation for the purpose of simulating certain chemical processes. One such alternative is adiabatic quantum computation~\cite{FarhiGoldstoneGutmannSipser}. This method of quantum information processing is naturally suited to problems which relate to ground states of complicated Hamiltonians, for example, the problem of protein folding~\cite{PRA78012320}.

The central mechanism of adiabatic quantum computation is the transformation of a prepared state to a final, desired state, by the gradual alteration of the system Hamiltonian. The algorithm requires a Hamiltonian $\left( \hat H_A \right)$ whose ground state is easily found, and a Hamiltonian $\left( \hat H_P \right)$ whose ground state encodes the solution to an appropriate mathematical problem. The adiabatic algorithm consists of three steps: prepare the ground state of  $\hat H_A $, vary the Hamiltonian according to
\begin{equation}
\hat H(t) = \left( 1-\frac{t}{\tau}\right)\hat H_A + \frac{t}{\tau}\hat H_P,  
\end{equation}
with $\tau$ sufficiently large so that the process is approximately adiabatic, and perform an appropriate measurement on the final state, which encodes the solution to the problem. 

This method can be straightforwardly applied to energy minimization problems, such as protein folding. Perdomo, et al.~\cite{PRA78012320} produced an instance of the adiabatic algorithm which is designed to find low-energy configurations for lattice-based hydrophobic-polar protein models. While it is unknown whether this model can be solved for arbitrary cases by any computer, quantum or classical, it is reasonable to conclude that a quantum algorithm would be of use, as the process of protein folding is inherently quantum-mechanical~\cite{Luo1,Luo2}. 

The hydrophobic-polar model of protein structure~\cite{dill1985} is a simplified model of protein folding, taking the individual peptides to be either hydrophobic (H), or polar (P). These peptides are distributed in a chain on a 2D or 3D lattice, and an energy is assigned to the configuration of the chain by counting the number of adjacencies between P peptides. This energy assignment based on spatial co-ordinates is mapped to a Hamiltonian consisting of three terms; an on-site term which penalizes conformations that have multiple peptides on the same site, a pairwise interaction term that rewards assignments with adjacent hydrophobic peptides, and a structure constraint Hamiltonian that penalizes spatial conformations that do not form chains. Since the positions of the peptides are the quantities of interest, a measurement in the computational basis is sufficient; no advanced measurement technique is required. 

We have seen that quantum information processing methods can be of great benefit to theoretical chemistry, providing properties of ground states, as well as reaction parameters. In the next section, we describe in further detail the requirements for a quantum computer. This is followed by a report on the current state of QIP implementations using nuclear and electronic spins in an external magnetic field. These NMR and ESR experimental approaches are well-suited to digital QIP, but not adiabatic methods, so we will focus on digital methods in the remaining sections. 

\subsection{NMR and the DiVincenzo criteria}
In order to perform one of the simulation algorithms mentioned above, there must first be a system which displays the unique properties of quantum mechanics, while allowing experimental control and measurement. This idea has been formalized into five well-accepted requirements known as the DiVincenzo criteria~\cite{DVC}: (1) a scalable physical system with well-characterized qubits, (2) the ability to initialize the register to a simple fiducial state, such as $\ket{0}^{\otimes n}$, (3) a universal set of quantum gates,(4) a qubit-specific measurement capability, and (5) decoherence times much longer than the average gate operation time. These criteria are partially satisfied by NMR QIP, where algorithms are implemented on nuclear spins in molecules, subjected to a large, constant external magnetic field and radio-frequency (RF) pulses ~\cite{vandersypen}. This system has the advantage of being well-tested arena for the development of quantum control techniques in the few-qubit regime (up to 12 qubits as of this writing \cite{PhysRevLett.96.170501}). Below, we describe the relationship between the desired properties of a quantum information processor and the particular features of the NMR system. 

\subsubsection{Scalability with Well-Characterized Qubits: \\ Spin-1/2 Nuclei}

In order for a quantum information scheme (an algorithm or a processor) to be scalable, the resources it requires must grow only polynomially with the size of the system.  Any implementation of QIP using $k$ two-level subsystems (referred to as qubits) is scalable in the total energy of the system, or equivalently, the precision with which the energy is measured.  NMR QIP employs such an architecture, as each spin-1/2 nucleus has two well-defined energy levels in a magnetic field, described by a two-dimensional Hamiltonian,
\begin{equation}
\hat{H} = \frac{1}{2}(\hbar \gamma B_0 +\delta) \hat Z = \frac{1}{2}(\hbar \gamma B_0 +\delta)  \left[ \begin{array}{cc}
1&0\\0&-1
\end{array}\right],
\end{equation}
where $B_0$ is the external magnetic field, $\gamma$ is the nuclear gyromagnetic ratio, and $\delta$ is the chemical shift term, imposed on each nucleus by its local molecular environment.  When nuclei have distinct $\delta$, they provide qubits that can easily be individually addressed through RF pulses. If this is not the case, more advanced techniques are required to address the qubits. 
 

NMR spectroscopy has been widely used in analytic chemistry and other disciplines for decades before quantum information research began, resulting in a large body of literature dedicated to measuring and modelling the NMR spectra for a given molecule~\cite{Abragam,spindynamics}. The techniques developed in conventional NMR can be used to determine the spin Hamiltonian, and thus characterize qubits, to a sufficient level of precision for quantum information. However, the number of qubits in NMR QIP systems is constrained in a practical sense, because the chemical shifts do not grow with the number of qubits, limiting the available frequency space for qubit addressing. Using solid-state NMR techniques, it is possible to use a magnetic field gradient to render the qubits distinguishable without the need for distinct $\delta$. However, it is difficult in practice to achieve gradients large enough to provide frequency resolution for nearby spins. 

Liquid-state NMR experiments have successfully demonstrated universal control over NMR systems ranging from a few qubits in the late 1990s and early 2000s~\cite{PhysRevLett.81.2152,vandersypen_experimental_2001} to up to twelve qubits more recently~\cite{PhysRevLett.96.170501}. While this number is not expected to appreciate greatly, NMR QIP experiments with up to $\sim 20$ qubits are likely achievable. The few-qubit experiments we have today  are sufficient for testing small quantum simulations and general concepts of QIP. Perhaps more importantly, the techniques developed in liquid- and solid-state NMR QIP are exportable to other, more scalable quantum information architectures, on which simulations of the type discussed in this chapter that go beyond the capability of the best classical computers could be performed.

\subsubsection{Initialization: The Pseudopure State}
\label{sec:initialization}
To perform a quantum algorithm, we need the register to begin in a known state such as  $\ket{0}^{\otimes n}$. In NMR, the Boltzmann distribution for a single spin has a bias toward the ground state of $\alpha \sim \tanh \left(\nicefrac{\hbar \gamma B_0}{k_BT}\right)$.  In ideal conditions, it would be possible to extract a pure state in the liquid state from a multi-qubit Boltzmann distribution. However, this is difficult in practice. Instead, it is possible to produce a multi-qubit \textit{pseudopure} initial state,
\begin{equation}
\rho_{\textrm{pseudopure}}=\frac{1-\alpha}{2^n} \hat I + \alpha \left(\ket{0} \! \bra{0} \right)^{\otimes n}.
\end{equation}
For typical values of $\gamma$ and $B_0$, $\alpha$ is on the order $10^{-5}$. Unitary operations due to applied pulses will alter the $\left(\ket{0}\!\bra{0} \right)^{\otimes n}$ term \emph{just as they would a pure state}, while leaving the identity term unchanged (neither is the identity term measurable, as it does not contribute to the total spin magnetization). As long as $\alpha$ is large enough to produce a measurable NMR signal, the final pseudopure state can be measured. 

As the size of the system increases, the methods that have been used to produce pseudopure initial states will result in values of $\alpha$ that decay exponentially in the system size, precluding scalability in liquid state NMR QIP. However, experimental procedures such as dynamic nuclear polarization and algorithmic cooling, to be discussed in Section \ref{sec:electronuke}, could increase nuclear polarizations to values near unity in suitable solid state systems that include electron spins. 

\subsubsection{A Universal Set of Quantum Gates: RF Pulses and Spin Coupling}
It is possible to implement an arbitrary $n$-qubit unitary gate using only one- and two-qubit gates~\cite{PhysRevA.51.1015}.  Single-qubit gates are realized in NMR QIP by exerting a time-dependent, radiofrequency (RF) magnetic field over the sample in the $\hat X - \hat Y$ plane (the $\hat Z$ direction being determined by the background static magnetic field), which implements the rotation operators $\hat U_x (\alpha)$ and $\hat{U}_y(\beta)$ (or a rotation along any axis in the $\hat X - \hat Y$ plane of the Bloch sphere) , where $\alpha$ and $\beta$ are arbitrary angles determined by the strength and duration of the RF pulses. Placing unitaries of this type in sequence can produce any possible single-qubit operation. Conditional logic (also known as \textit{if-then} logic) in NMR QIP is achieved by allowing the nuclear magnetic state to evolve under coupling mechanisms present in the molecule. These coupling mechanisms are direct dipolar coupling and electron-mediated dipolar coupling (i.e. indirect dipolar coupling ), the latter also known as $J$-coupling. 

One can also express the universality of an implementation of QIP directly in terms of the Hamiltonian, in contrast to examining the unitary operators generated by the Hamiltonian. This is more useful for quantum simulations, since the class of Hamiltonians which can be simulated by a physical system is easily obtained from the form of the natural and control Hamiltonians~\cite{AskIain}. Universal control over a system, in this sense, is the ability to simulate any Hamiltonian with the same number of degrees of freedom. Importantly, any coupled Hamiltonian, together with universal single-qubit control, can be used to simulate any other coupled Hamiltonian that has the same degree of connectivity \cite{AskIain}. 

Since all of the spin-$\frac{1}{2}$ nuclei in a molecule possess magnetic moments, they will possess a pairwise direct dipolar coupling term in the NMR Hamiltonian:
\begin{equation}
\hat{H}^{DD}_{jk}=-\frac{\mu_0}{4 \pi} \frac{\hbar \gamma_j \gamma_k}{r_{jk}^3} (3(\vec \sigma_j \cdot \vec{e}_{jk})(\vec \sigma_k \cdot \vec{e}_{jk})-\vec{\sigma_j} \cdot \vec{\sigma_k}),
\end{equation}
where $r_{jk}$ is the distance between the two nuclei, $\vec{\sigma}=[\hat X,\,\hat Y,\,\hat Z]$ and $\vec{e}_{jk}$ is a unit vector along the line connecting the two nuclei. In the liquid state, $\Theta_{jk}$ (the angle between $\vec e_{jk}$ and the external $\vec B$-field axis) is, to a good approximation, uniformly distributed by fast molecular tumbling. This results in zero average interaction strength. Therefore, direct dipolar coupling is only useful in systems in which the $\Theta_{jk}$ are not uniformly distributed, such as solid state NMR (where all of the $\Theta_{jk}$ are fixed by  crystal orientation) and liquid crystal NMR (where the $\Theta_{jk}$ are distributed non-uniformly). In addition to the direct dipolar coupling, there exists an indirect coupling, known as the $J$-coupling. This is a relatively weak coupling mediated by the electron cloud in the molecule, according to 
\begin{equation}
\hat H^{J}_{jk} =  2 \pi \vec \sigma_j \cdot \hat J_{jk} \cdot \vec \sigma_k^{\intercal} ,
\end{equation}
where $\hat J_{jk}$ is a $3 \times 3$ real matrix. This coupling term does not vanish, even under rapid tumbling. In isotropic liquids, it has the Heisenberg form $\vec \sigma_j \cdot \vec \sigma_k$, which reduces to  $\hat Z_j  \hat Z_k$ when the two spins have a chemical shift difference much larger than the value of $J$; this is the typical generator of a two-qubit gate in liquid-state NMR.  The variety of natural coupling terms renders QIP possible on many different molecules, in the liquid, solid, or liquid crystalline states.  

\subsubsection{Measurement: Free Induction Decay}
The apparatus (an RF coil) used to implement rotations about axes in the $\hat X - \hat Y$ plane can also be used to detect ensemble magnetization signals from the sample in the $\hat X - \hat Y$ plane. In order to translate the logical state of an NMR QIP register into an $\hat X$ observable value, a readout pulse of angle $\frac{\pi}{2}$ is applied about the $\hat X$-axis at each qubit resonance frequency. This results in a signal, proportional to the pseudopure parameter $\alpha$ (see Subsection \ref{sec:initialization}), from which the logical state can be extracted. 
This process does not result in projective measurement, but still allows universal computation including full state tomography~\cite{SekharCory2004}. In addition, the expectation value of any Hermitian operator can be obtained in a single shot, allowing the measurement of arbitrary operators by quadrature detection~\cite{spindynamics}.


\begin{table}
\begin{center}
\begin{tabular}{cccccc}
\cline{1-6}
&$C_1$(Hz)& $C_2$(Hz)& $C_m$(Hz)& $T_2^{\ast}$(ms)& $T_1$(s)\\
\cline{1-6}
$C_1$ & 5693 & 237 & 828 & 2.4 & 162  \\
$C_2$ & & 1748 & 1020 & 2.6 & 326  \\
$C_m$ & & & -3358 & 3.1 & 314 \\
\cline{1-6}
\end{tabular} 
\caption{ (from~\cite{osamamastersthesis}) Characteristic timescales for control in solid-state NMR QIP, using the $^{13}$C-labelled malonic acid molecule, can be gleaned from the dipolar coupling strengths (off-diagonal elements in second and third columns) and chemical shifts (diagonal elements), and compared with relaxation and dephasing times (rightmost columns). Note the description of decoherence in terms of $T_2^{\ast}$, which combines decoherence from inhomogeneous and homogeneous sources. Much longer $T_2 \approx 100$ ms was observed with suitable dipolar refocusing / dynamical decoupling sequences applied \cite{baughSSNMR}.}
\label{table:omttable}
\end{center}
\end{table}

\subsubsection{Noise and decoherence: \\ $T_1 / T_2$ Vs. Coupling Strength}
In order to perform an algorithm, the time used to implement the appropriate pulses must be much shorter than the characteristic timescale of decoherence. In addition to the  noise related to imperfect implementation of the gates (these are typically `coherent' errors), the unavoidable interaction with the surrounding environment leads to true decoherence. In NMR, there are two main mechanisms for this decoherence: thermal equilibration (energy relaxation) and dephasing, characterized by timescales $T_1$ and $T_2^{\ast}$, respectively. $T_2^{\ast}$ represents the ensemble, inhomogeneous dephasing time of the system, whereas $T_2$ is the intrinsic decoherence time of each single quantum subsystem. $T_2^{\ast}\leq T_2$ can often be improved to $T_2$ by applying decoupling pulses as described in Section \ref{sec:DD}. In NMR QIP, the coupling terms between qubits are usually the smallest terms in the Hamiltonian, leading to relatively slow two-qubit gates. The solid-state malonic acid system provides an example, shown in Table \ref{table:omttable}; the carbon-carbon coupling ranges from 200 Hz to 1 kHz. This indicates that the associated controlled-$\hat Z$ gate requires a time of $1/2J \sim $ 0.5 - 2.5 ms, allowing the implementation of several two-qubit gates before a single decay period $T_2^{\ast}$ has elapsed. This suffices for simple algorithms, but a larger ratio of decoherence-to-gate time is required in order to implement more complex algorithms, and in particular, to implement error correction. If the noise is below a threshold~\cite{Nielsen2000}, it is possible to use fault-tolerant methods to counteract its effects. If the noise exceeds this threshold, control must be improved until the threshold is attained.  It is possible to assess the noise level using benchmarking~\cite{PhysRevA.77.012307,doi:10.1139/p07-192}.  In an NMR system where control was optimized, it has been possible to reach an error per gate rate of $10^{-4} $ for single-qubit gates and $10^{-3}$ for two-qubit gates~\cite{1367-2630-11-1-013034}. 

Ongoing improvements to the implementation of one- and two-qubit gates in magnetic resonance QIP have arisen, in part, due to: 
\begin{itemize}
\item optimization of RF pulses to produce high-fidelity unitary operations,
\item isolating the system of interest from unwanted degrees of freedom, and
\item taking advantage of the unique properties of the electron spin. 
\end{itemize}
In the following sections, we describe recent progress in these directions.
\section{Quantum Control in Magnetic Resonance QIP}
\subsection{Advances in Pulse Engineering}
\label{sec:PE}
In NMR QIP, pulse engineering is the practice of developing control techniques for either generating coherent transfer from an initial spin state to a desired spin state (state-to-state transfer) or producing an effective Hamiltonian that implements a desired unitary gate $\hat U$ (unitary propagator) by manipulating the external RF field. This can be achieved by augmenting the internal Hamiltonian with a control Hamiltonian (in the laboratory frame), 
$$\hat{H}_{c}(t)=\summation{k}{}\frac{\omega_{k}(t)}{2}[\cos(\omega_{\mathrm{RF}}t+\phi_{k}(t))\hat X+\sin(\omega_{\mathrm{RF}}t+\phi_{k}(t)) \hat Y],$$
where $\omega_{k}(t)$ and $\phi_{k}(t)$ denote the amplitude and the phase applied at a transmitter frequency $\omega_{\mathrm{RF}}$.The time evolution is typically divided into a set of $N$ steps. At each step $j$, the evolution is given by the unitary propagator $\hat{U}_{j}=\text{exp}\{-i\Delta t \hat{H}_{j}(t)\}$, where $\hat{H}_{j}(t)$ is the total Hamiltonian with a control set $\{\omega_{k}^{j},\ \phi_{k}^{j}\}$. The final density operator is given by $\rho(T)=\hat{U}_{N}...\hat{U}_{1}\rho_{0}\hat{U}_{1}^{\dagger}...\hat{U}_{N}^{\dagger}$.  The task of pulse engineering is to find a set of $\{\omega_{k}^{j},\ \phi_{k}^{j}\}$ so that the resulting dynamics are sufficiently close to those desired. The quality of the pulse is evaluated by a fidelity function $\Phi$ that is proportional to overlap between the obtained operator and the target operator.

Pulse engineering techniques must address experimental limitations for practical applications: qubit selectivity due to finite frequency bandwidth, instrumental errors (such as static and RF field inhomogeneity),  miscalibration of pulse power or duration, and frequency offset. 

Pulse design must take these artefacts into account in order to achieve robust pulses with high fidelity on the ensemble. This causes a substantial increase in the complexity of the pulse design problem.  For example, errors caused by chemical shift dispersion, RF inhomogeneity and RF miscalibration can be suppressed if the parameters are sampled over a range of discrete values determined by the uncertainty. Then the total fidelity function to be optimized can be measured as $\Phi_{\text{tot}}=\summation{\alpha}{}\Phi(x_{\alpha})$, where $x_{\alpha}$ is a particular value of some parameter seen by a fraction of spins, e.g. RF field amplitude. In this case, the optimization problem becomes significantly more difficult, since the effective parameter space is much larger. 

Over the last 30 years, numerous techniques have been developed for control in NMR QIP. Traditionally, average Hamiltonian theory (AHT) has  been a powerful tool that provides intuitive guidelines for constructing pulse sequences in relatively simple cases~\cite{waugh}. In particular, composite pulses~\cite{composite}, adiabatic pulses~\cite{adiabetic1, adiabetic2, adiabetic3, adiabetic4}, and shaped pulses~\cite{shaped} were introduced during the earlier development of NMR QIP to better compensate for static and RF field inhomogeneity by increasing the number of degrees of freedom in the pulse shape. However, the long pulse times produced by these techniques lead to greater decoherence and relaxation effects, and interference of selective pulses simultaneously applied to different spins. However, not all errors can be corrected using these techniques~\cite{str_mod}. Strongly modulated pulses~\cite{str_mod} average out unwanted evolution by using strong control fields to drive the spins, so that a desired unitary operator can be synthesized directly. This method uses high-power pulses that decrease the required pulse duration, and hence reduce the effect of decoherence and relaxation. Penalty functions on power, frequency and time period can be applied to obtain more experimentally acceptable solutions. Moreover, it allows for the incorporation of robustness against errors caused by static or RF field inhomogeneities. Strongly modulated pulses with ideal fidelities of order $~99\%$ are performed for universal control of liquid-state NMR system with up to six qubits~\cite{str_mod}. The method was also studied in a three-qubit solid-state NMR system, single crystal malonic acid. Universal control with ideal fidelities $~98\%$ was achieved in the strong-coupling regime of the direct nuclear dipole-dipole couplings, in which the state evolutions under the internal Hamiltonian are more complex than those of typical liquid-state NMR system~\cite{baughSSNMR}. In addition, the transformations generated by the strongly modulated pulses were shown to be robust against the ensemble inhomogeneities that are present in solid-state NMR systems~\cite{baughSSNMR}. The cost of this method (as well as the optimal control methods described below) is that the full system dynamics must be simulated in the course of finding pulses, so that even using clever methods such as gradient-based approaches to improve efficiency does not render the method scalable to arbitrary numbers of qubits. 

Recently, optimal control theory (OCT), originally developed for problems in engineering~\cite{krotov_ref1,krotov_ref2,krotov}, has been used for systematic optimization of pulse designs in NMR QIP. Analytical solutions to time-optimal realization of unitary operations can be obtained by formulating a variational principle to reduce the problem to a set of first-order ordinary differential equations~\cite{timeoptimal1, timeoptimal2}. For more than two qubits, the analytical approach is intractable, and numerical methods are required. 

\begin{figure}[h!]
\centering
\includegraphics[width=0.36\textwidth]{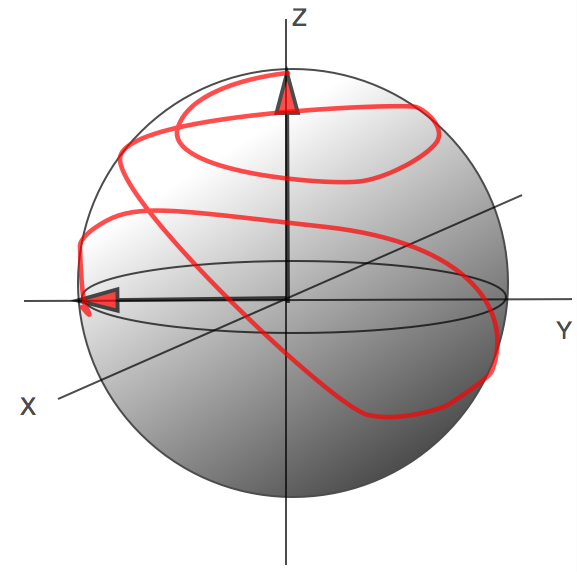}
\caption{Trajectory of the spin magnetization vector on the Bloch sphere for a pulse found by the GRAPE method~\cite{grape}. The pulse is designed for a $\frac{\pi}{2}$ rotation around the $\hat{X}$-axis.}\label{fig:grape_trajectory}
\end{figure}

For larger systems, the majority of control methods for optimizing the efficiency of coherent transfer are based on a gradient approach, such as the Gradient Ascent Pulse Engineering (GRAPE) algorithm~\cite{grape}. The GRAPE algorithm can be summarized as follows. It begins with an initial guess for a set of control parameters $u_{k}(j)$ for all time steps $j \le N$. Then in each iteration the algorithm evaluates $\delta\Phi_{tot}/\delta u_{k}(j)$ for $j \le N$ and updates $u_{k}(j)$ as: $$u_{k}(j)\rightarrow u_{k}(j)+\epsilon\frac{\delta\Phi_{tot}}{\delta u_{k}(j)}.$$ The iteration continues until the improvement in the performance index $\Phi_{tot}$ is smaller than a chosen threshold value. GRAPE also evaluates penalty functions to reflect realistic experimental constraints. The gradient approach enables GRAPE algorithm to calculate full time evolutions much faster than conventional (brute force) numerical difference methods. Consequently, the number of pulse parameters to be optimized can be orders of magnitude larger than in conventional approaches.
A number of recent NMR QIP experiments~\cite{jingfuQEC, PhysRevLett.100.140501, jpolynomial, gina_qdiscord, osama_contextuality, MSdistill, 2roundsQEC} have utilized GRAPE pulses to achieve high control fidelities. However, just as with any optimal control method, the GRAPE algorithm cannot deterministically locate global minima but instead finds local minima in the search space. Therefore the ultimate fidelities that can be achieved are limited by the initial guess, even if unit fidelity pulses are possible in principle. Figure~\ref{fig:grape_trajectory} illustrates an example of a path taken by the spin magnetization vector on the Bloch sphere during a GRAPE pulse designed for a $\frac{\pi}{2}$ rotation around the $\hat{X}$-axis.
 
Another well-developed numerical method for quantum control is Krotov-based numerical method~\cite{krotov_origin, Tannor, Zhu, Maday}. The Krotov-based method allows large changes in control parameters from one iteration to the next, and immediately exploits all available information at each time step, monotonically improving the objective fidelity functional at each iteration. Maximov \textit{et al.}~\cite{krotov} analysed the Krotov-based algorithm in the context of NMR spectroscopy and compared it to GRAPE, concluding that the Krotov-based algorithm consumes less computational resources per iteration and is much better than GRAPE for making an initial guess towards a global maximum.  However, the larger step size in the Krotov-based method limits efficiency as the solution gets closer to the desired fidelity.  There is an open question as to whether high efficiency can be obtained by combining the two methods, using the Krotov algorithm to quickly prepare a good initial pulse sequence to load into GRAPE for final refinement~\cite{krotov}.

As mentioned above, a major drawback with these methods is that computations cost grows quickly with increased system size.  As quantum processors become larger, new techniques will need to be developed that might allow these OCT methods to be applied within blocks of nearby qubits (in frequency space) and scalable pulse design techniques between blocks \cite{ryan2008}.

\subsection{Advances in Dynamical Decoupling}
\label{sec:DD}

In order to achieve high-fidelity control in NMR QIP, one must address system-environment interaction, as well as unintended evolution under internal couplings between spins. The coupling interaction is ``always on'' and therefore must be suppressed to perform single qubit gates.  This process is called refocusing and is relatively simple in liquid state NMR, since the only coupling term is of the form $\hat Z_j \hat Z_k$.  Solid-state NMR Hamiltonians have more complex coupling terms, and are therefore more difficult to refocus.  Techniques based on average Hamiltonian theory, such as magic angle spinning~\cite{MAS}, Lee-Goldburg decoupling~\cite{LGD} and multiple-pulse techniques \cite{waugh}, have proven useful in refocusing these terms. Furthermore, recent developments in optimal control theory (see Section~\ref{sec:PE}) allow the refocusing of unwanted internal interactions for arbitrary coupling forms, as long as the Hamiltonian is well-known. 

Hereinafter, we narrow the discussion to system-environment, or dynamical decoupling (DD). A DD scheme is a sequence of control fields applied to a system for some time with the objective of increasing coherence time by attenuating the system-environment interaction. Starting from Hahn's discovery of spin-echo in 1950~\cite{spinechoes}, many methods to accomplish this have been actively studied, one seminal example being the Carr-Purcell-Merboom-Gill (CPMG) sequence~\cite{CPMG} which is successful at suppressing single-axis, low-frequency noise. The performance of realistic DD sequences is limited by experimental imperfections (see Section~\ref{sec:PE}), so it is important to design a DD scheme that is more robust to the most relevant errors.

In traditional DD, a pulse sequence is periodically applied to reduce undesired terms of the system-bath interaction Hamiltonian, known as periodic dynamical decoupling (PDD). These sequences were improved in 2005 by Khodjasteh and Lidar~\cite{FTDD}, who introduced concatenated DD pulse sequences (CDD), such as $p_{n+1} = p_{n}\hat X p_{n}\hat Zp_{n}\hat Xp_{n}\hat Z$, where $p_{n}=\tau_{0}$ is a delay between pulses and $n$ is a concatenation level. CDD is significantly more efficient at decoupling than PDD with equal pulse numbers and is more robust to both random and systematic control errors. CDD can reduce the system-environment interaction to order $t^{n+1}$ where $t$ is total duration of the cycle and $n$ is the concatenation level. The caveat of this method is that due to the nature of concatenation the number of necessary pulses grows exponentially ($4^{n}$) in the concatenation level, whereas order of suppressed error only grows linearly. 

In 2007, Uhrig introduced an optimized dynamical decoupling sequence based on gradient moment nulling in NMR~\cite{UDD_origin}, called Uhrig Dynamical Decoupling (UDD)~\cite{Uhrig, UDD1, UDD2}. UDD was first introduced for a specific spin-boson model, and was later discovered~\cite{UDD1, UDD2} to be system independent. What distinguishes UDD from conventional DD schemes is that the time delay blocks between $\pi$ pulses are not equal. Uhrig showed that splitting the total time interval $t$ into smaller intervals $0, \delta_{1}t, ..., \delta_{n}t$ (where $\delta_{j}=\sin^{2}[\pi j/(2n+2)]$), can suppress decoherence up to order $t^{n+1}$ without exponential cost. Moreover, Uhrig's simulation showed that the performance of decoupling from the environment becomes independent of system-bath coupling strengths for long sequences. One drawback of UDD is that it works for only single-axis error, so for example, cannot simultaneously suppress $T_{1}$ and $T_{2}$ processes. 

In an attempt to refocus the effects of noise about multiple axes, Uhrig concatenated the UDD sequence (CUDD)~\cite{CUDD}.  The total number of pulses to suppress decoherence and relaxation to order $t^{n}$ is proportional to $2^{n}$, decreasing the resources by $2^n$ from CDD sequences. West \textit{et al.} further improved on this by creating a new UDD-based sequence that suppresses decoherence and relaxation to order $n$ using $(n+1)^{2}$ pulse intervals~\cite{west}. Their scheme integrates two sequences, $T_{1}$-correcting UDD and $T_{2}$-correcting UDD, and is known as quadratic UDD (QDD).

UDD-based methods are optimal when the noise has a sharp high-frequency cutoff~\cite{uhrig_NJP, pasini, UDD1, noise_spec1, noise_spec2}. However, in a low-frequency-dominated noise environment, conventional CP (or CPMG) is preferred~\cite{pasini, noise_spec1, noise_spec2}. Borneman \textit{et al.}~\cite{troy} pointed out that, although the CPMG sequence is inherently tolerant of field inhomogeneities and pulse calibration errors, the robustness of the sequence is limited by the quality of the RF pulses used. They adapted the GRAPE algorithm (see Section~\ref{sec:PE}) and applied iterative optimization method to design  general refocusing pulses (universal $\pi$ rotation around the $\hat Y$-axis) with high fidelity over a wide range of resonance offset frequencies and RF amplitudes. Borneman et al were able to find a pulse which refocuses $99\%$ of the initial magnetization over a range of frequency offsets of $\pm 10$ KHz (four times greater than the maximum RF amplitude) for uniform RF field. For RF inhomogeneity of $\pm 10\%$, they were able to find a pulse that refocuses $98\%$ of the initial magnetization over a frequency range of $\pm 8$ KHz (3.2 times the maximum RF amplitude), an improvement over any previously published refocusing pulses of similar duration and maximum RF amplitude when applied in a CPMG sequence.

Souza \textit{et al.}~\cite{souza} compared the performance of a number of pulse sequences in order to find a decoupling scheme that is robust against pulse imperfections or control errors. They considered two possible approaches to compensate the imperfections: use composite pulses that the error correction is done \ql inside" the pulse (robust pulse), or create the sequence where the error from one pulse is averaged out in subsequent pulses (self-correcting sequence). In the experiment a standard pulse (non-robust) and the Knill pulse~\cite{NV} $\pi_{\pi/6+\phi}-\pi_{\phi}-\pi_{\pi/2+\phi}-\pi_{\phi}-\pi_{\pi/6+\phi}$, which is robust against flip-angle and off-resonance errors are inserted into the CPMG, periodic XY-4 \cite{xy-4}, and CDD. Then decay times of the magnetization as a function of the duty cycle, the total pulsing time in the sequence divided by total duration of the sequence, are analysed and compared for each case. A new sequence called Knill Dynamical Decoupling (KDD) which is achieved by inserting delays between the $\pi$ pulses of the Knill pulse, is also examined. For small duty cycles, sequences with non-robust pulses are superior, due to the shorter cycle time of the non-robust pulse sequence when constant duty cycles are compared. However, the performance of these sequences saturates or decreases with increasing duty cycle. Comparatively, robust pulses continue to increase in performance for large duty cycles. KDD is found to have the best performance for large duty cycles, and comparable to sequences without robust pulses for small duty cycles.

\subsection{Control in the Electron-Nuclear System}
\label{sec:electronuke}
Many of the techniques used to control spin-$\nicefrac{1}{2}$ nuclei in NMR can also be applied to electrons, comprising electron spin resonance (ESR). As the $\gamma$ of an electron is $\approx 660$ times greater than that of a proton, this leads to a much larger Larmor frequency and higher polarization, but also faster decoherence and relaxation than nuclear states. The coupling between electron and nuclear spins is governed by the hyperfine interaction: 
\begin{equation}
\hat H_{\mathrm{hf}} = \vec \sigma_E \cdot \mathcal{A} \cdot \vec \sigma_N^{\intercal},
\end{equation}
where $E$ denotes the electron, $N$ denotes the nucleus, and $\mathcal{A}$ denotes the hyperfine coupling tensor. The combined electron-nuclear solid state spin system is appealing for QIP because it is possible to exploit the strengths of each type of spin: electron spins possess higher polarization for initialization and readout, and can be manipulated on faster timescales, while nuclear spins are ideal for long-term storage of coherent states~\cite{MZhangESR}.

Although ESR provides much better signal-to-noise ratio than NMR, it usually suffers from broader linewidths. For an electron-nuclear spin system, one is required to characterize its internal Hamiltonian as precisely as possible in order to achieve high fidelity control. However, the nuclear Zeeman term in the Hamiltonian is small relative to terms involving the electron, and often it is not possible to measure directly in an ESR experiment due to spin selection rules~\cite{PrinciplesOfEPR}. Moreover, the ESR linewidth can be comparable to the hyperfine splitting which in turn makes it difficult to study the hyperfine interaction term. Fortunately, the well-established magnetic resonance techniques of electron spin echo envelope modulation (ESEEM)~\cite{PrinciplesOfEPR, ESEEM} and electron nuclear double resonance (ENDOR) spectroscopy~\cite{PrinciplesOfEPR, cwENDOR, pulsedENDOR} can precisely quantify nuclear transition frequencies and hyperfine couplings. In ENDOR, both NMR and ESR transitions are driven directly. Thus ENDOR spectroscopy is a promising technique for controlling electron-nuclear systems for QIP. However, it should be noted that ENDOR experiments require additional RF hardware, and nuclear spin flip times are limited by the small nuclear gyromagnetic ratio and the strength of the RF field that can be achieved. 
 
The anisotropy of hyperfine interaction allows controlling nuclear spins solely by irradiating electron spin transitions~\cite{PhysRevA.78.010303}. The indirect control technique is advantageous: it simplifies the instrument design as an additional RF interface is not needed, and faster control of a nuclear spin is achievable when the hyperfine coupling strength exceeds the Larmor frequency of the nucleus at given external field~\cite{khaneja_switched_2007}. The following section reviews this approach and recent experiments demonstrating the universal control in electron-nuclear systems. 
\subsubsection{Indirect Control via the Anisotropic Hyperfine Interaction}
\begin{figure}[h!t]
\centering
\includegraphics[width=0.85\textwidth]{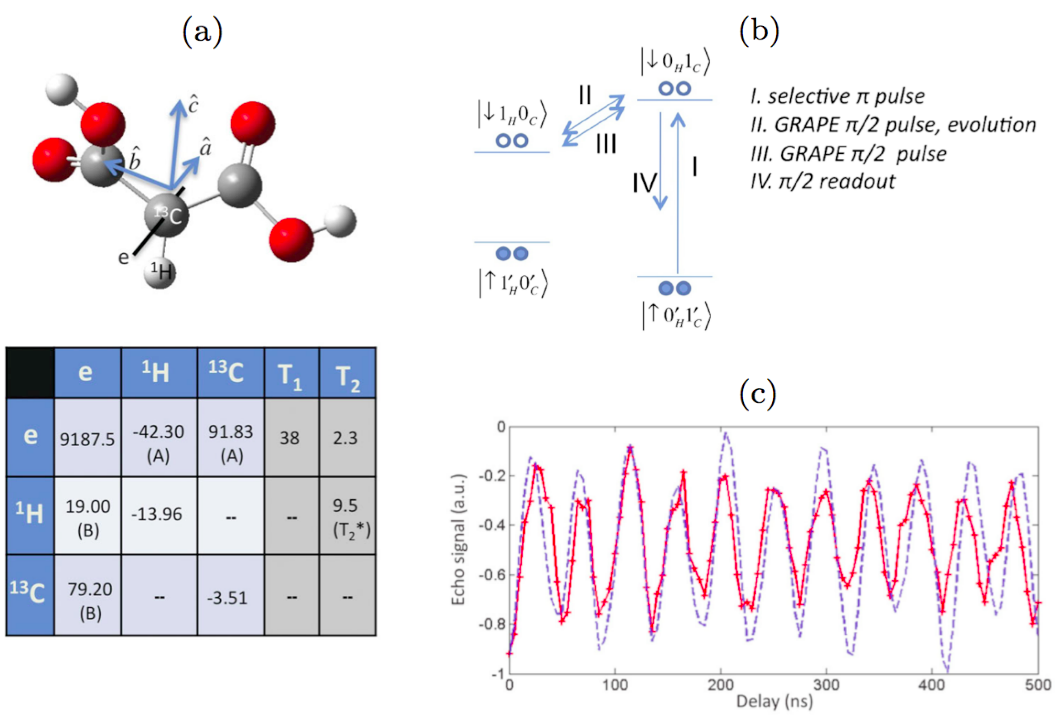}
\caption{From~\cite{MZhangESR}. (a) upper: Schematic of the singly $^{13}$C-labelled malonic acid radical. lower: Table of Hamiltonian parameters for the crystal orientation used in the experiment. Larmor frequencies are along the diagonal, and hyperfine coupling coefficients are off-diagonal, with all frequencies in MHz. $T_{1}$ and $T_{2}$  are shown in the two right-most columns in microseconds. (b) Schematic of the experimental sequence for creating and detecting a double-nuclear coherence in the electron spin down manifold. (c) The double-nuclear coherence echo signal versus delay time in experiment (red crosses, solid line) and simulation (blue dashed line) for the transition between $\kket{\downarrow 1_{H}0_{C}}$ and $\kket{\downarrow 0_{H}1_{C}}$ in (b) (electron spin-down manifold).}\label{fig:MZhang}
\end{figure}

If the coupling frequency due to hyperfine interaction is larger than the nuclear Larmor frequency, efficient nuclear control can be obtained by manipulating the electron~\cite{khaneja_switched_2007}, producing an effective magnetic fields seen by the nucleus that depends on the electron spin state. With the anisotropic coupling, the nucleus sees an effective field
\begin{equation}
\vec B_{\mathrm{eff}} = \left( B_0 \pm \dfrac{A}{2\gamma_N} \right) \hat z \pm \dfrac{B}{2\gamma_N} \hat x,
\end{equation}
where $A$ is the $zz$-component of the hyperfine tensor \cite{PrinciplesOfEPR} and $B$ is $\sqrt{\mathcal{A}_{zx}^2+\mathcal{A}_{zy}^2}$, the component of the hyperfine tensor in the $\hat{x}$-direction on the nucleus, in a specially chosen frame. $B_0$ is the static magnetic field and $\gamma_N$ is the nuclear gyromagnetic ratio. The $\pm$ sign takes the value $+$ when the electron spin is parallel to the external field, and $-$ when it is anti-parallel. Any rotation of the Bloch sphere can be generated by repeated rotations about these two distinct axes. Universal single-qubit control of the nuclear spin can then be achieved indirectly, since the free precession of the nucleus about these axes does not require RF pulses resonant with nuclear transitions. Hodges \textit{et al.} implemented GRAPE algorithm (see Section~\ref{sec:PE}) and showed universal control of a 1e-1n system (malonic acid radical)~\cite{PhysRevA.78.010303}. In~\cite{MZhangESR}, Zhang \textit{et al.} also utilized the GRAPE method to perform an entangling gate between two nuclear spins  in a 1e-2n system (singly $^{13}$C-labelled malonic acid radical). Schematics of the malonic acid radical, an energy level diagram showing creation and detection of double-nuclear coherence, and the double-nuclear coherence echo signal from~\cite{MZhangESR} are shown in Figure~\ref{fig:MZhang}. These experiments represented the first steps toward reaching high-fidelity universal control of a one-electron, N-nuclear spin hyperfine coupled system by using the electron as an actuator.

In the following sections, we discuss additional quantum information techniques which exploit the properties of combined NMR/ESR systems:
\begin{itemize}
\item dynamic nuclear polarization and algorithmic cooling, which increase the polarizations of nuclear spins,
\item spin-bus implementations and parallel information transfer, which produce two-qubit gates between nuclear spins mediated by electrons.
\end{itemize}
These two techniques, described below, reinforce the potential of combining nuclear and electron spin control to better satisfy the DiVincenzo criteria in NMR QIP. 
\subsubsection{Dynamic Nuclear Polarization and Algorithmic Cooling}
A nuclear spin in a magnetic field at room temperature has an equilibrium density operator described by a Boltzmann distribution, with a typical bias of $\sim 10^{-5}$. Although this is sufficient to demonstrate the principles of QIP, it does not allow for resetting ancilla qubits as needed, for example, in quantum error correction. Since the nuclear magnetic moment is so weak, this problem cannot be simply resolved by brute-force cooling or increasing the magnetic field. For example, reducing the temperature to 1 K and using the largest available magnetic fields increases the polarization only to the order $10^{-3}$. Therefore, it is necessary to develop novel techniques to provide initial polarization that approaches unity.  

The much larger electron magnetic moment can be used as a polarization source. A \textsc{swap} gate between a nuclear spin and an electron, each at thermal equilibrium, acts as a polarization swap and gives the nuclear spin a large non-equilibrium polarization. This type of process is called \textit{dynamic nuclear polarization}, or DNP. In the context of QIP, such polarization transfers have been implemented in $^{15}$N@C$_{60}$~\cite{morley_efficient_2007}, achieving a single-spin nuclear polarization of 62\% at 4.2 K, in an 8.6 T field. Also, DNP has been used in conjunction with spin diffusion in silicon microparticles to produce bulk spin polarizations up to 5\% at 1.5 K and 2.35 T~\cite{dementyev_dynamic_2008}. Nuclear $T_1$ for these systems was shown to be dependent on the particle size, making the polarization decay controllable in principle. The $^{15}$N nucleus at a Nitrogen vacancy (NV) centre has been initialized to a polarization of $98\%$ by optically pumping the NV defect \cite{PhysRevLett.102.057403}, and similar polarizations can be reached for proximate $^{13}$C nuclei. These examples show that control of a single electron spin coupled to an NMR QIP register can be used to great advantage.  

It is possible to obtain nuclear qubits with still greater polarization by an implementation-independent method, algorithmic cooling. This technique combines reversible entropy compression (unitary process) and open quantum systems cooling (non-unitary process). For an m-qubit system with each qubit possessing equal entropy, a reversible entropy compression process can compress entropy into $\text{m}-\text{n}$ qubits, increasing their effective spin temperature while cooling the subset of n qubits. Here, lower entropy corresponds to higher polarization. The system is in contact with a large heat reservoir (bath) whose entropy is lower than the initial entropy of each system qubit. SWAP gates are then applied to exchange the system entropy and the heat bath entropy to cool down $\text{m}-\text{n}$ qubits that were heated during the compression step. The local part of the heat bath quickly relaxes back to its equilibrium state and reacquires its initial high polarization. This two-step process is applied iteratively until no further cooling is possible. Given m qubits and a heat bath polarization of $ \epsilon_b > 2^{-m} $, it is possible to cool the system very close to its ground state with only polynomially increasing resources~\cite{PhysRevLett.94.120501}. However, for $ \epsilon_b < 2^{-m} $, the maximum polarization attainable by a single spin is $\epsilon_b 2^{m-2}$~\cite{osamamastersthesis}. Algorithmic cooling has recently been implemented in liquid and solid state NMR~\cite{2005quant.ph.11156B, PhysRevLett.100.140501, 2011arXiv1108.5109E}. Since electron spins possess higher polarization and much have shorter $T_1$ times than nuclear spins at a given magnetic field strength and temperature, having nuclear qubits able to interact with an electron spin `bath' is a promising path to nuclear qubit initialization. Single quantum systems, such as NV centres with a sufficient number of coupled $^{13}$C spins, would be ideal for demonstrating nuclear qubit purification by algorithmic cooling. 

\subsubsection{Spin Buses and Parallel Information Transfer}
Since a single electron can couple to multiple nuclei, it can be used to transfer information between them, creating an effective coupling. This indirect coupling is the basis of the S-bus~\cite{mehring_spin-bus_2006}, useful for performing multi-qubit gates when the electron-nuclear and electron-electron coupling are much stronger than the nuclear-nuclear coupling. This concept was used to perform Deutsch's Algorithm~\cite{mehring_spin-bus_2006}, using a system of two nuclear qubits coupled by an electron in CaF$_2$:Ce$^{3+}$. 

\begin{figure}[h!t]
\centering
\includegraphics[width=0.6\textwidth]{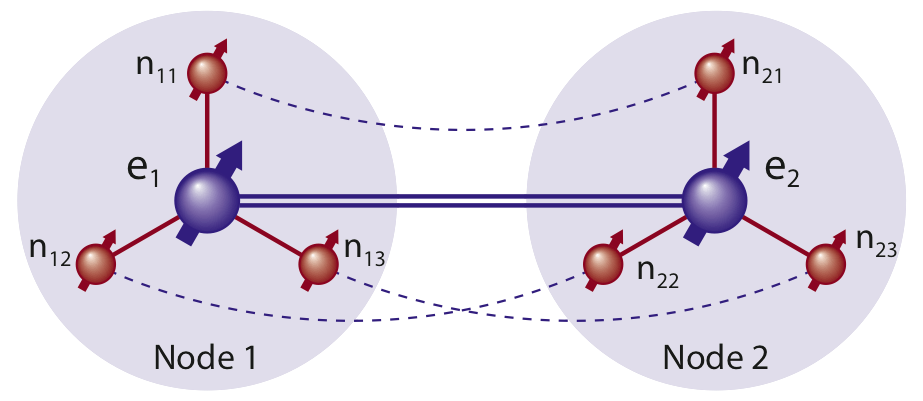}
\caption{Schematic of 2 x (1e-3n) Node: The nodes are taken to be identical, with resolved anisotropic hyperfine interactions (solid red lines) between electron actuator spins and nuclear processor spins. The local processors are initially disjoint, but may be effectively coupled (dotted lines) by modulating an isotropic actuator exchange or dipolar interaction (solid blue double line) and moving into an appropriate microwave Hamiltonian interaction frame. The spin labelling is $e_{i}$ for electron actuator spins and $n_{ij}$ for nuclear processor spins, where $i$ and $j$ label the nodes and the qubits, respectively.~\cite{2011arXiv1107.4333B}}\label{fig:actuator}
\end{figure}

The indirect coupling of nuclei using information transfer of the type described above has two flaws which must be overcome to confirm its utility as a nuclear control method. The first is that the state being transferred is subject to strong decoherence and relaxation while it is stored on the electron. The second is that only two nuclei can be coupled through the bus at the same time; two-qubit gates cannot be performed in parallel. These problems were recently resolved~\cite{2011arXiv1107.4333B}, where the assumed architecture consists of two local nodes (taken to be sets of $n$ nuclei, each coupled to an electron via anisotropic hyperfine interaction), with the only inter-node coupling (dipolar or exchange interaction) being between electrons. By using the interaction frame, Borneman, Granade and Cory showed that states of the multi-nucleus nodes can be swapped in parallel, effectively performing $n$ two-qubit gates simultaneously. Also, the effect of decoherence on the electron is mitigated by ensuring that no computational state is stored on the electron. Figure~\ref{fig:actuator} illustrates the idea presented in~\cite{2011arXiv1107.4333B}.

The electron spin, when used as a component of the NMR QIP system, possesses properties that complement those of the nuclear spins. While the nuclear spin has a longer coherence time and low initial polarization, the electron has a short coherence time and high initial polarization. The electron-nuclear coupling also permits an array of techniques, making the electron a valuable asset to NMR QIP. 

An example of electron-nuclear system application for QIP that has been extensively investigated recently is entanglement generation~\cite{e1/2-n1/2Ent, e3/2-n1/2EntPRL, e3/2-n1/2EntJCP, RobabehThesis, B819556K, SimmonsEnt}. Preparing entangled states in NMR QIP is impracticable since required spin polarization according to the positive partial transpose (PPT) criterion~\cite{PPT1, PPT2} for entanglement is far above what is reachable. Electron-nuclear QIP at high field and low temperature, in conjunction with optimal control techniques, can potentially provide a highly entangled quantum state that is a key ingredient in many quantum algorithms and applications.

\section{NMR QIP For Chemistry}
\subsection{Recent Experiments in NMR Quantum Simulation}
The advances described in the previous section can enable the development of larger, more precise processors for quantum information. We focus now on three recent experiments that display the current capabilities of the NMR QIP system, and prove its utility as a testbed for future implementations. These experiments are motivated by the study of physical systems rather than chemical processes, however, the methods within are easily adapted to problems of chemical interest. 
\subsubsection{Simulation of Burgers' Equation}
One challenge which presents itself to scientists of many disciplines is the solution of non-linear differential equations. Numerical solution of these equations is costly, this cost having motivated the development of numerous simulation methods. One such method, which takes advantage of quantum resources, is the quantum lattice gas algorithm on a type-II quantum computer. 

\begin{figure}[h!t]
\centering
\includegraphics[scale=0.3]{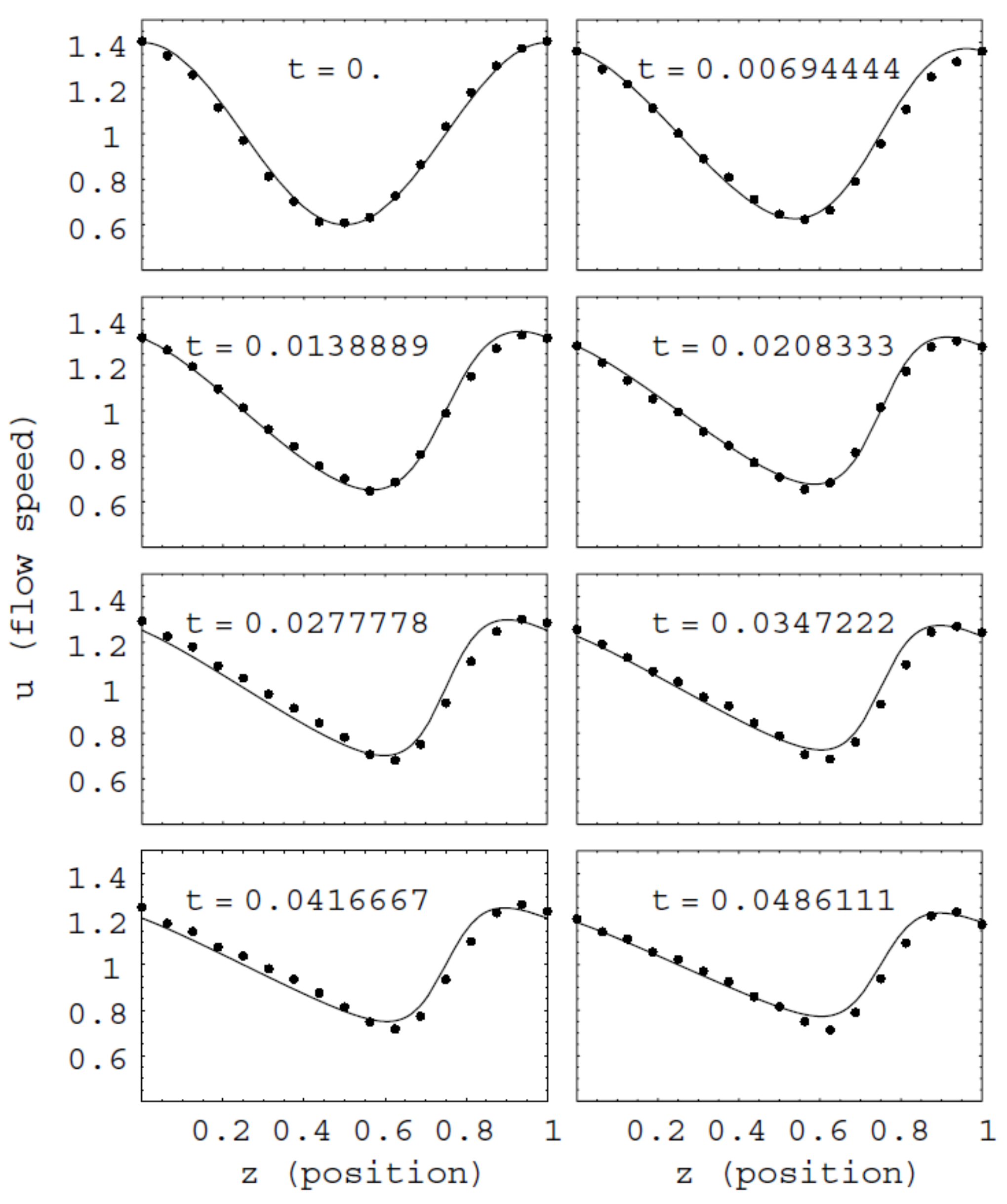}
\caption{From~\cite{CoryBurgers}. Comparison of analytic and numerical solutions to Burgers' equation. Numerical data has been obtained on a 16-node Type-II quantum computer.}\label{fig:CoryBurgers}
\end{figure}

Type-II quantum computers~\cite{YepezPRA66012310} are networks of small quantum processors which communicate classically to form a large analog computer. Operations that can be carried out in parallel are well-suited to implementation on such a processor, since the individual QIP nodes are distinct and can therefore be addressed simultaneously. The quantum lattice gas algorithm consists of such operations; it is divided into three steps which are iterated:
\begin{enumerate}
\item A state is prepared on the individual QIP nodes, corresponding to initial (or transient) conditions of the dynamical model being studied. The state on a given node is 
\begin{equation}
(\sqrt{1-f_a(x_l,t_0)}\ket{0}+\sqrt{f_a(x_l,t_0)}\ket{1})^{\otimes 2}
\end{equation}
denoting two copies of the state encoding the value of the function to be propagated. 
\item A \textit{collision operator} $\hat U$ is applied in parallel across all of the nodes, which simulates a single finite time step for the dynamics being modelled.  
\item A measurement is made on the nodes and the resulting transient state is fed back to step 1. Classical communication is used, in this step, to transfer the results of step 2 between QIP nodes. 
\end{enumerate}
The resulting class of finite difference equations which can be numerically solved by this method is that of the form
\begin{equation}
f_a (x_{l+e_a},t_{n+1})=f_a(x_l,t_n)+ \bra{\psi (x_l,t_n)} \hat U^{\dagger} \hat n_a \hat U - \hat n_a\ket{\psi (x_l,t_n)}
\end{equation}
where $e_a=\pm 1$, $\hat U$ is the collision operator, and $f_a$ is the function whose evolution is being simulated. 

Discretization of Burgers' equation yields a finite difference equation of the form described above, it is a non-linear differential equation which describes shock formation in fluid dynamics:
\begin{equation}
\partial_t u(z,t) + u(z,t) \partial_z u(z,t) = \nu \partial_{zz}u(z,t)
\end{equation}
Due to the simplicity of this equation, it can be solved analytically. From~\cite{CoryBurgers}, the analytical solution to Burgers' equation is compared to the simulation on a 16-node type-II NMR quantum processor, shown in figure \ref{fig:CoryBurgers}. 

\begin{figure}[h!t]
\centering
\includegraphics[scale=0.35]{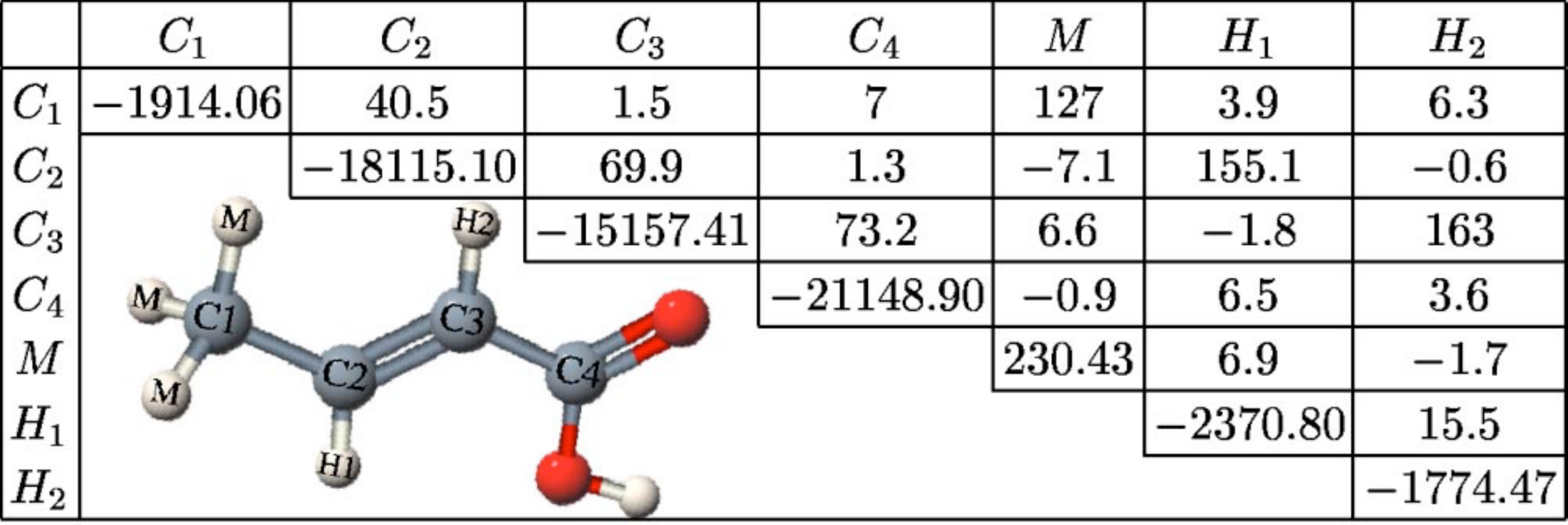}
\caption{From~\cite{Camille}. Spin Hamiltonian parameters and molecular diagram for transcrotonic acid.}\label{fig:Camille1}
\end{figure}

\subsubsection{Simulation of the Fano-Anderson Model}
Another challenge with which physicists are often presented is that of simulating the evolution under a Hamiltonian representing the energy landscape of a large number of identical particles. The Fano-Anderson model possesses such a Hamiltonian, acting on $n$ spin-less fermions constrained to a ring, surrounding an impurity:
\begin{equation}
\hat H = \sum_{l=0}^{n-1}\epsilon_{k_l}c^{\dagger}_{k_l}c_{k_l} + \epsilon b^{\dagger}b+ V(c^{\dagger}_{k_0}b+b^{\dagger}c_{k_0}),
\end{equation}
where $c_{k_l}$ is a fermionic annihilation operator on the conduction mode $k_l$, $b$ is a fermionic annihilation operator on the impurity in the center, $c_{k_l}^{\dagger}$ and $b^{\dagger}$ are the respective creation operators, and $\epsilon_{k_l}$, $\epsilon$ and $V$ denote the strengths of the terms in the Hamiltonian corresponding to occupation of a site, occupation of the impurity and tunnelling between sites and the impurity, respectively. This Hamiltonian is exactly diagonalizable, however, its simulation is an important step in validating QIP methods for application to simulations in condensed matter theory.

\begin{figure}[h!t]
\centering
\includegraphics[scale=0.3]{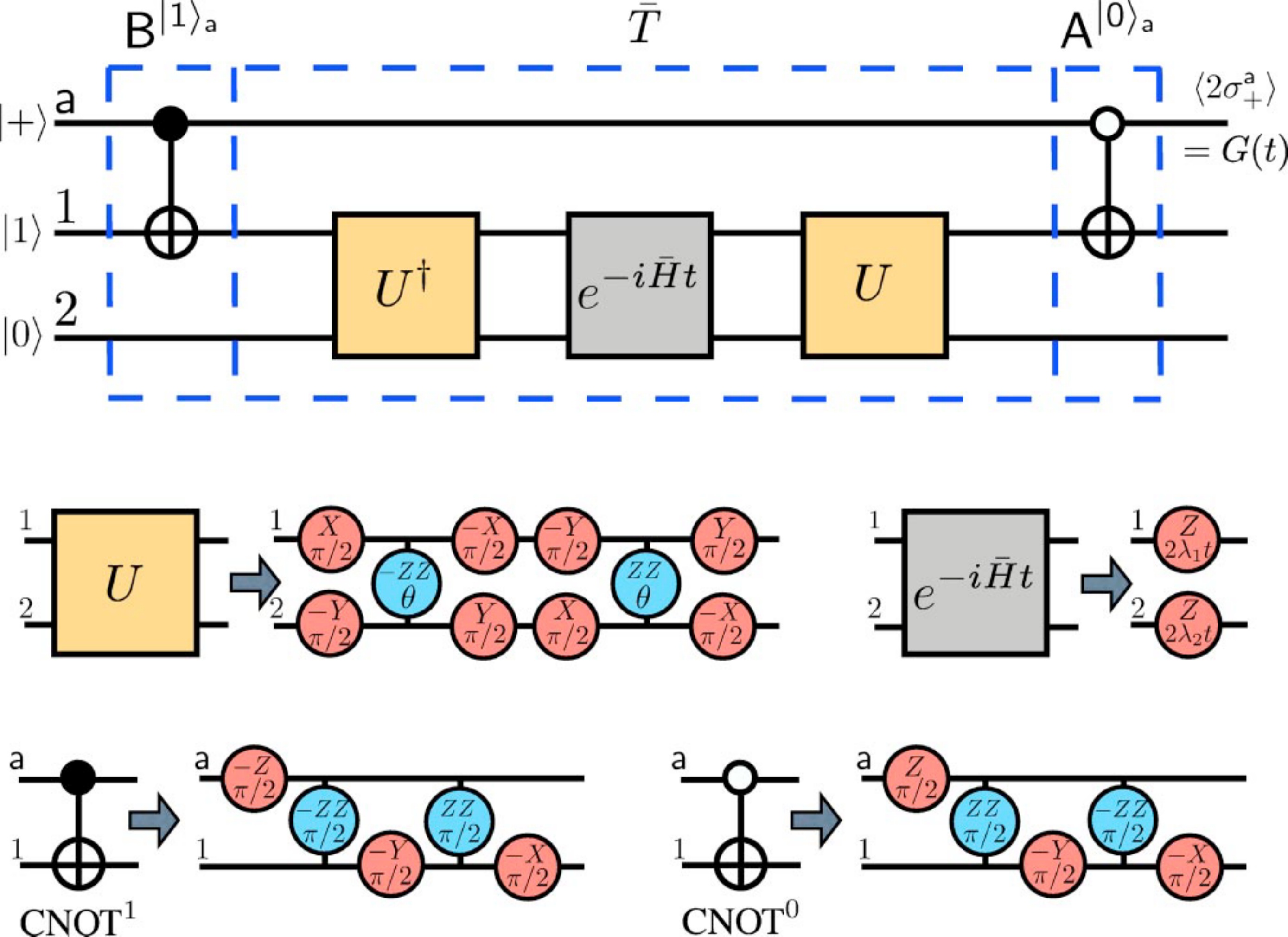}
\caption{From~\cite{Camille}. The kernel of the quantum simulation is the implementation of a unitary operator that evolves the qubit states according to the desired Hamiltonian. Given the capability to produce controlled operations between an ancilla qubit and the simulation register, correlation functions of interest can be directly measured.}\label{fig:Camille2}
\end{figure}

The simulation performed by Negrevergne, et al.~\cite{Camille} concerns the smallest possible Fano-Anderson model, one site interacting with an impurity. This algorithm requires three qubits, one for the site, one for the impurity and one ancilla qubit, required for indirect measurement. It is implemented in a liquid-state NMR system, using transcrotonic acid, whose Hamiltonian parameters are shown in Figure \ref{fig:Camille1}. 

Again, the simulation is divided into three broad steps. However, iteration is unnecessary in this experiment, since the Hamiltonian to be simulated can be exponentiated continuously. First, the system is prepared in an initial state, based on the pseudopure state. Next, in order to perform an indirect measurement, a two-qubit gate couples the ancilla to the simulator, which is then evolved according to the Fano-Anderson Hamiltonian, and decoupled using another two-qubit gate, shown in Figure~\ref{fig:Camille2}.

\begin{figure}[h!t]
\centering
\includegraphics[scale=0.25]{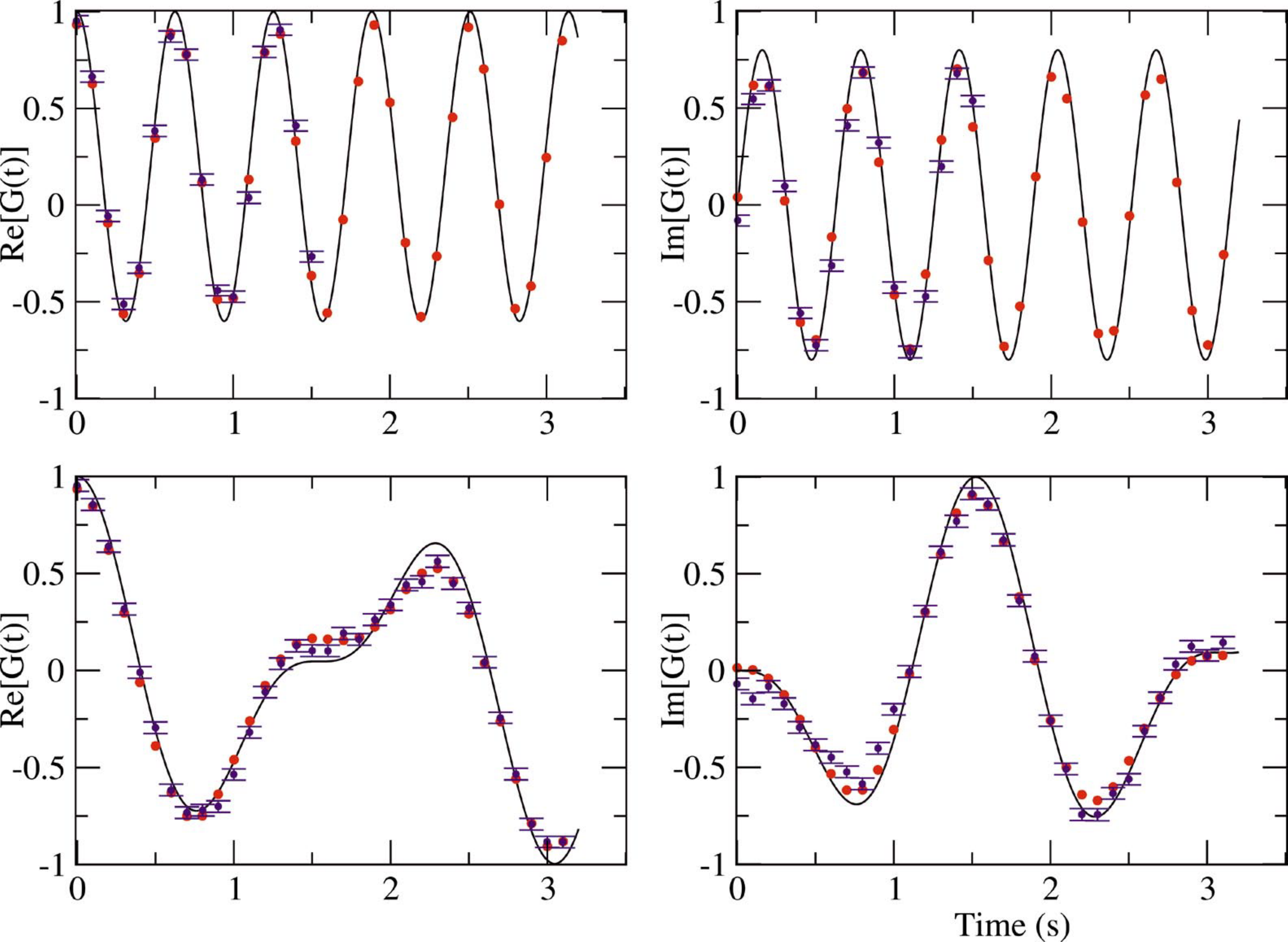}
\caption{From~\cite{Camille}. Analytic and numerical solutions for the correlation function $G(t)$, whose evaluation is given diagrammatically in Figure \ref{fig:Camille2}.}\label{fig:Camille3}
\end{figure}

Two quantities of interest were obtained in this experiment. $G(t)$, the correlation between the states $b^{\dagger}(t)\ket{FS}$ and $b^{\dagger}(0)\ket{FS}$, where $\ket{FS}$ is the state corresponding to the filled Fermi sea. In the NMR QIP system, this correlation function is $\bra{10} \exp \left(i \bar H t \right) \sigma_x^1 \exp \left(-i \bar H t \right) \sigma_x^1 \ket{10}$. The results of the measurement of this function in the simulation are shown in Figure \ref{fig:Camille3}.
Another quantity, of more general interest, is the spectrum of the Hamiltonian. This can be measured with a simpler quantum circuit, and results in excellent agreement with theory were also obtained in reference~\cite{Camille}. 

\begin{figure}[h!]
\centering
\includegraphics[scale=0.75]{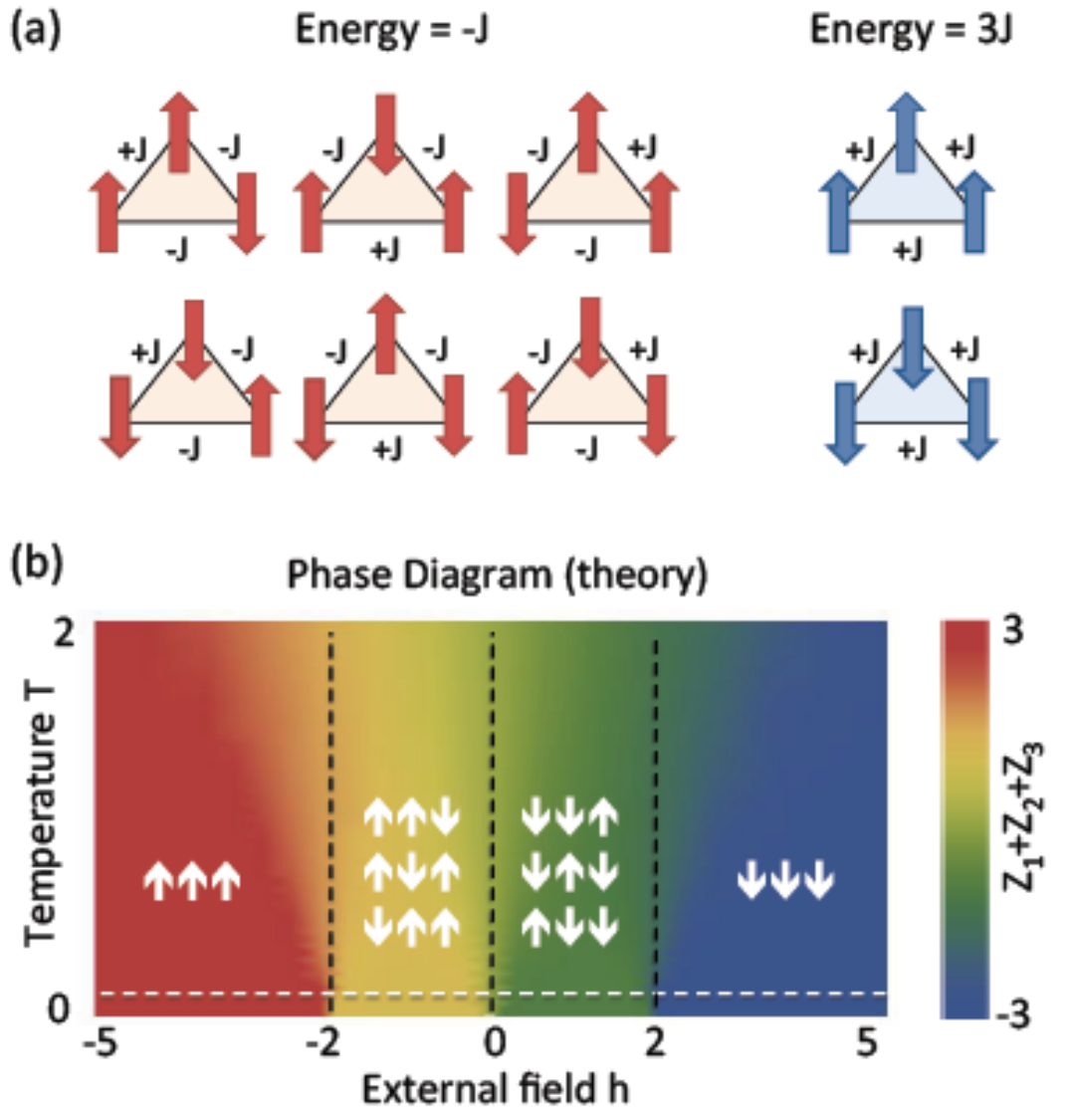}
\caption{From~\cite{Jingfu}. (a) Spin configurations of the Ising Hamiltonian at zero temperature and field $h$. There is a six-fold
degeneracy in the ground state, leading to a non-zero entropy. (b) Thermal phase diagram for the same Hamiltonian showing total magnetization versus field and temperature.}\label{fig:Jingfu1}
\end{figure}

\subsubsection{Simulation of Frustrated Magnetism}
\begin{figure}[h!]
\centering
\includegraphics[scale=0.25]{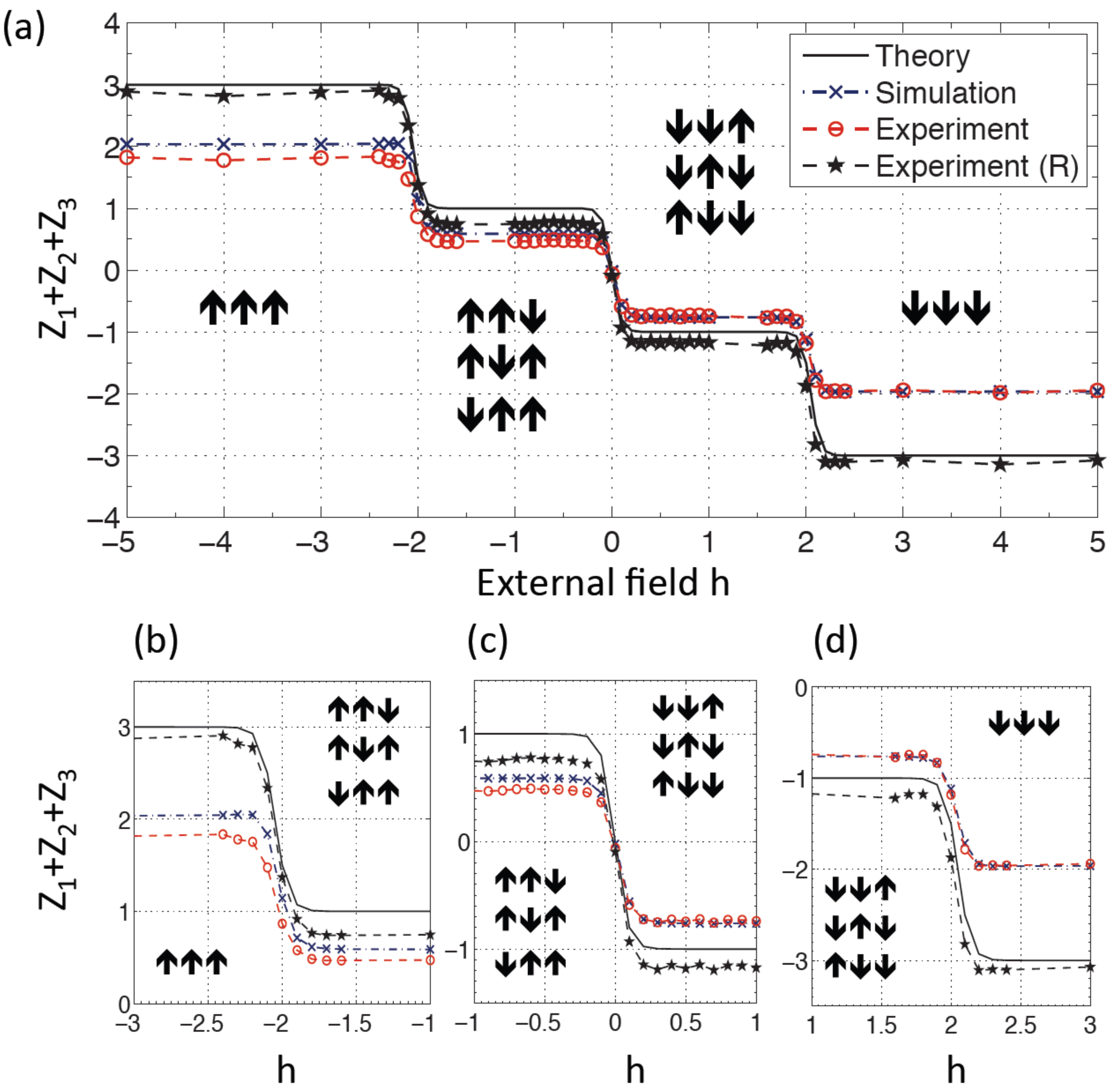}\\
\includegraphics[scale=0.35]{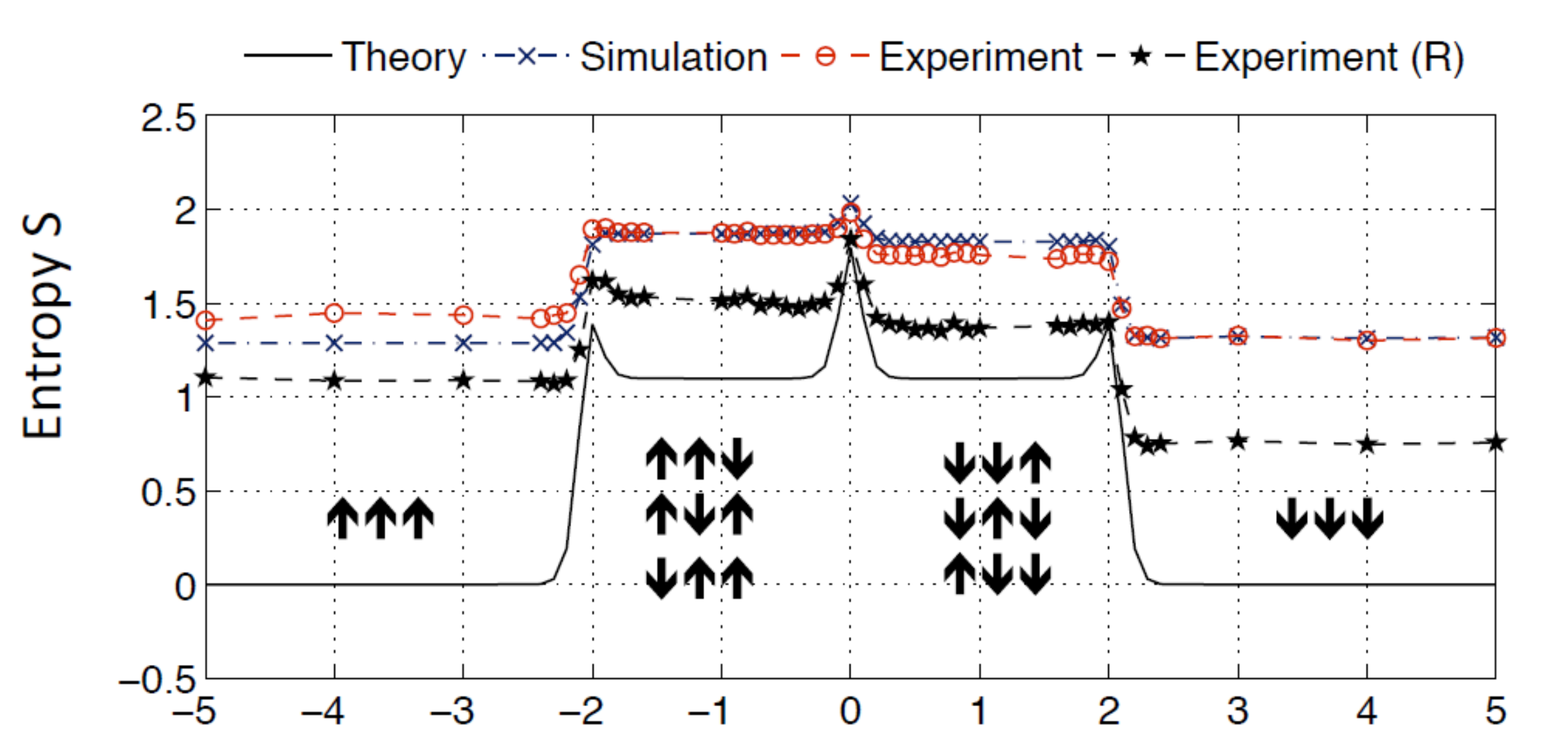}
\caption{From~\cite{Jingfu}. (a-d) The average magnetization of the three-spin anti-ferromagnetic Ising system versus simulated field $h$ at low temperature, showing theoretical predication, numerical simulation of the NMR experiment including decoherence, and experimental results. Below, entropy $S$ versus field $h$, also at low temperature. The entropy peak at $h=0$ indicates frustration.}\label{fig:Jingfu2}
\end{figure}

Any spin system which has interaction Hamiltonians that cannot be simultaneously minimized is geometrically frustrated. Frustration is a fundamental problem in the study of magnetism, since its presence means that the Hamiltonian cannot be divided into small subsystems which can then be studied individually to obtain information regarding a global property. This exacerbates the central problem of spin system simulation, i.e. the exponential  number of degrees of freedom required to simulate such a system. Zhang et al.~\cite{Jingfu} have performed a digital quantum simulation of the fundamental building block of such a frustrated system: the three-spin anti-ferromagnetic Ising model:
\begin{equation}
\hat H = J(\hat Z_1\hat Z_2+\hat Z_2\hat Z_3+\hat Z_1\hat Z_3) + h(\hat Z_1+\hat Z_2+\hat Z_3), 
\end{equation}
where $J>0$. This simple Hamiltonian can be solved analytically, due to its small size. 
The central technique in this work is to simulate an arbitrary \emph{thermal} state using pseudopure state preparation. In this manner, the phase diagram in $h$, $J$ and $T$ can be explored, deriving multiple properties of the resulting states. The pseudopure portion of the prepared state is 
\begin{equation}
\ket{\psi_{\beta}} = \sum_k \sqrt{\exp \left( -\beta E_k \right)/\mathcal{Z}}\ket{\phi_k}
\end{equation}
where the $\ket{\phi_k}$ are the eigenstates of the Hamiltonian with energy $E_k$, $\mathcal{Z}$ is the partition function and $\beta=1/T$. The two physical quantities which Zhang, et al. focus on are the total magnetization of the system $\hat Z_1+\hat Z_2+\hat Z_3$, and the entropy of the resulting state $S=\mbox{Tr} (\rho_{\beta} \log \rho_{\beta})$, where $\rho_{\beta}$ is the thermal density matrix of the Ising system (see Figure~\ref{fig:Jingfu2}). A peak in entropy is observed at $h=0$ indicating frustration of the magnet, in agreement with theory. 

These three experiments serve to demonstrate the validity of the principles of quantum simulation, and demonstrate the necessary control to achieve scientific results, in principle. The greatest challenge to implementation of these algorithms on a larger scale is the inability to control very large Hilbert spaces. The advances presented in section~\ref{sec:electronuke}, for example, may be helpful in obtaining this control. 

\section{Prospects for Engineered Spin-Based QIPs}
\label{sec:prospects}

Nuclear and electron magnetic resonance on bulk materials containing natural spin systems has been an excellent ground for testing ideas of quantum information processing  in the few-qubit regime. The progression to non-trivial numbers of qubits will most likely be accomplished by transitioning to engineered single spin systems, and much progress has taken place in this direction recently. Promising candidates include quantum dot electron spins \cite{Ciorga2001}, Nitrogen-vacancy centers in diamond \cite{Steiner2010}, donor electrons/nuclei in Si \cite{Kane1998}, and Nitrogen atoms trapped in C60 \cite{Morton2006}, among others. A common theme among many of these approaches is to use the electron spin for initialization, fast gate operations and readout, and to use nuclear spins for long-time qubit storage or, in the case of quantum dots, as a controllable local effective magnetic field \cite{Lai2006} (the nuclear spin in III-V quantum dots present an unfortunate decoherence problem when left uncontrolled \cite{Khaetskii2002}). Single qubit control is realized by some form of magnetic resonance in each of these approaches (with the exception of the singlet-triplet qubit in quantum dots \cite{Petta2005}), whereas two-qubit coupling strategies vary widely. Since these are single quantum systems rather than ensembles, spatial addressing or a mix of spatial and frequency addressing becomes possible, an important advantage for scalability over frequency-only addressing. These systems are in principle more readily scalable than bulk magnetic resonance of molecules, however, developing reliable qubits with single quantum systems typically carries a host of technological challenges. Though none of the approaches listed above has yet moved beyond a small handful of qubits, increasingly higher quality single and few-qubit systems are being realized at a rapid pace; some examples include demonstration of dynamical decoupling \cite{Ryan2010} and multi-qubit control \cite{Jelezko2004} in NV centers, single-spin readout in Si \cite{Stegner2006}, and high-fidelity single qubit control and refocussing in quantum dots \cite{Bluhm2011, Nadj-Perge2011}. 

A significant challenge is to achieve fast, high-fidelity readout of a single spin (note that projective readout of the spin state is quite different from simply detecting the presence of a magnetic moment). So far, the most promising methods are optical (e.g. spin-dependent optical transition in NV center) or via electron transport (e.g. using the Pauli spin blockade in a double quantum dot \cite{Petta2005,Nakul2008}). Atomic force \cite{Rugar2004} and nano-magnetometry methods \cite{Grinolds2011} are able to detect single spins, but require orders of magnitude improvement in sensitivity$/\sqrt{Hz}$ before single-shot spin readout becomes feasible. These magnetometry experiments have nonetheless opened up a wide range of possible applications in quantum sensing, i.e. exploiting quantum coherence to surpass classical limits on measurement sensitivity \cite{Grinolds2011}, and this technology may play an eventual role in spatial readout of spin-based quantum processors. See \cite{Jones2009} for a seminal demonstration of quantum sensing of magnetic fields using liquid-state NMR.   

Another common feature of these approaches is the solid-state environment surrounding the qubits, which typically leads to shorter decoherence times than more isolated systems like ion traps or liquid-state NMR. Although spin-1/2 particles are immune to direct coupling with electric fields (the dominant noise source in solids), the spin-orbit coupling together with phonons provides a pathway for spin relaxation of electrons, which can then in turn act as magnetic noise sources for nuclei. Here, dynamical decoupling cannot typically improve the situation, because the correlation time of the electron spin relaxation process is usually much shorter than the timescale of control. Low temperature is thus required in order to reduce the density of phonons and suppress relaxation, typically leading to electron spin relaxations times in milliseconds for defect centres in dielectric crystals and for spins in quantum dots \cite{Elzerman2004}, and up to minutes for electrons at shallow donors in high purity Si at 1.2K \cite{Tyryshkin2011}. The latter work in high purity Si demonstrated electron spin coherence times exceeding one second, a groundbreaking result for a solid-state system \cite{Tyryshkin2011}. 

A set of nuclear spins, evolving under  nuclear-nuclear dipole couplings, can also act a magnetic noise source for an electron coupled to one or more of them. This is the case for III-V quantum dots, where an electron  spin is coupled to $\sim 10^6$ nuclear spins at once due to the contact hyperfine interaction \cite{Coish2009}. The electron spin is dephased on a timescale $\sim 10$ns due to statistical fluctuations of nuclear polarization, but orders of magnitude longer dephasing times have been achieved with dynamical decoupling \cite{Bluhm2011}. Similar spin dynamics take place for the electron spin in NV centers which is coupled to a small number of proximate natural abundance $^{13}C$ nuclei, or for donor electrons in Si coupled to $^{29}$Si nuclei; again dynamical decoupling is seen to improve coherence times significantly \cite{Ryan2010, Morton2008} due to the slow correlation time of the nuclear bath. In the case of electronic defects in insulating or semiconducting crystals, the nearest `shell' of nuclei may have resolvable hyperfine couplings and can therefore be utilized as qubits, whereas more distant nuclei have unresolvable couplings and simply generate (dephasing) magnetic noise. 

Hybridizing spin and other quantum degrees of freedom, such as photons, may solve some of these challenges and is an active area of research. The aforementioned NV center is already a type of hybrid system in which electron spin can be readout and initialized optically, and furthermore, coherent quantum information stored in the spin state can be converted to a photonic `flying qubit' \cite{Fu2008}. The strong coupling cavity quantum electrodynamics regime has been demonstrated with NV center optical dipole transitions coupled to high finesse optical cavities \cite{Park2006, Englund2010}, opening the door to photon-mediated coupling of spatially distant spin qubits. Similar efforts are underway to achieve strong coupling between quantum dot spins and microwave photon modes in on-chip superconducting resonators, for example, using strong spin-orbit coupling to couple the spin qubit to the cavity electric field \cite{Trif2008}. Besides applications in quantum communication, this kind of approach could allow for distributing processing tasks optimally across different qubit realizations, and for minimizing the number of gates needed in a computation by increasing the effective dimensionality of the coupled network. 

Chemistry has played a role in the early development of quantum control in the context of bulk magnetic resonance of molecular ensembles, and one can envision `bottom-up' architectures emerging for scalable QIP based on patterned molecular arrays or monolayers, e.g. those investigated in the molecular electronics research area \cite{Green2007, Tour2000}. A viable approach would likely use some form of stable radical to provide electron spins for initialization, fast manipulation, and readout, and nuclear spins for ancillae and quantum memory. The principle challenges in such a system would be addressing and readout of spins on individual molecules. Addressing could be achieved using suitably strong magnetic field gradients to encode spatial information in the frequency domain, and with suitable improvements in sensitivity it might be possible to use a nano-diamond NV center scanning probe `read head' to achieve single-shot spatial readout of spin qubits \cite{Grinolds2011}. The techniques for electron-nuclear hyperfine control discussed in section~\ref{sec:electronuke} would then be invaluable tools for implementing a set of high fidelity, universal quantum gates in such a system. \\

\textbf{Acknowledgments}
We thank G. Passante for helpful discussions, and Y. Zhang for help in compiling references. The Natural Science and Engineering Research Council of Canada provided support for the writing of this document. 

\bibliographystyle{unsrt}
\bibliography{BookChapterReferences}
\end{document}